\documentclass[12pt,aps]{revtex4}   
\usepackage{amsmath}        
\usepackage{amssymb}        
\usepackage{amsbsy}         
\usepackage[mathscr]{eucal} 
\usepackage{graphicx}       

\begin{document}
\title{Band structure and Bloch states in birefringent 1D magnetophotonic crystals: An analytical approach}
\author{Miguel Levy, Amir A. Jalali}
\affiliation{Michigan Technological University, Department of Physics, 1400 Townsend Dr, Houghton, MI 49931, USA}


\bibliographystyle{prsty}

\begin{abstract}
An analytical formulation for the band structure and Bloch modes in elliptically birefringent magnetophotonic crystals is presented.  The model incorporates both the effects of gyrotropy and linear birefringence generally present in magneto-optic thin film devices.  Full analytical expressions are obtained for the dispersion relation and Bloch modes in a layered stack photonic crystal and their properties are analyzed.  It is shown that other models recently discussed in the literature are contained as special limiting cases of the formulation presented herein.\\
\end{abstract}


\maketitle

\section{Introduction}
Photonic crystals in magnetic systems have been the subject of a number of theoretical and experimental studies in recent years.\cite{Figotin2001,Fedyanin2004,Khanikaev2005,Baryshev2004,Inoue1999,Levy2001,Steel2000a,Levy2006a}  Interest in these systems stems from the  scientific and technological possibilities resulting from the combination of magnetic tunability, non-reciprocal effects such as Faraday rotation, and band gap engineering.\cite{Figotin2001,Baryshev2004,Inoue1999,Levy2001,Steel2000a,Levy2006a} Examples of interesting phenomena that have been investigated include electromagnetic unidirectionality, frozen light generation, the magnetic tuning of photonic bandgaps, and Faraday rotation enhancement.\cite{Baryshev2004,Inoue1999,Levy2001,Steel2000a,Levy2006a,Kahl2004a,Figotin1998,Figotin2003}

Work reported in the literature on such photonic crystals has dealt with systems in which at least one of the constitutive component materials possesses circular birefringence.\cite{Inoue1998}  These are called gyrotropic or bigyrotropic materials.  However, technological imperatives such as miniaturization, the development of integrated devices and optical waveguide fabrication demand that polarization effects resulting from stresses and optical confinement be taken into account.  Magneto-optic films grown by various deposition techniques and subject to lattice mismatch and differential thermal expansion with the substrate will acquire linear birefringence.\cite{Steel2000a,Yang2004,Huang2006}  In addition, in optical channels formed on the film, transverse electric (TE) and transverse magnetic (TM) modes will generally possess different effective or modal refractive indices, hence linear birefringence.\cite{comment}  The combination of circular and linear birefringence results in elliptically polarized modes, as pointed out, for example, in a recent publication by one of us.~\cite{Levy2006} Hence we prefer to use the term elliptical birefringence in this case.

These phenomena have an impact on the polarization state of light propagating in a magneto-optic photonic crystal and on its band structure.  Although some work has been done on the transmittance and polarization response in magnetic photonic crystal waveguides, the effect of elliptical birefringence on the band structure and Bloch modes in such systems has not been directly studied.\cite{Levy2006a,Li2005}  Recently Khanikaev and co-workers presented an analytic solution to the band structure of a photonic crystal model consisting of ultra-thin magneto-optic layers in a stack~\cite{Khanikaev2005}.  Circular birefringence was included in the model but not the elliptical birefringence that results from combined optical gyrotropy and stresses or optical confinement.  In the present article we present an analytic solution for the case of a gyrotropic photonic crystal stack with elliptical birefringence.  The model does not assume ultra-thin magneto-optic layers and contains the solution obtained by Khanikaev and co-workers as a special case in the limit of zero linear birefringence and ultra-thin magnetic layers. In addition, the model does not confine itself to quarter-wave plates but allows the layers to have arbitrary thickness limited only by the period of the crystal. The model supports the combination of magnetic and non-magnetic layers as well.

The value of an analytical solution to the case of birefringent magnetophotonic crystals is that it gives explicit form to the dependence of the band structure and the Bloch states on the material and geometrical parameters of the crystal.  It thus represents a useful tool in the analysis and design of photonic band gap systems.  The formulation presented here shows that the wavevector dependence of the optical frequencies is sensitive to the relative birefringence between adjacent layers in the photonic crystal stack.  And it gives explicit functional form to this dependence for Bloch states of different helicities.  Layer thickness and thickness ratios between different layers also strongly impact the band structure, as shown in the discussion on Bloch modes and the dispersion relation.

\section{Wave propagation in birefringent magnetophotonic media}
The wave equation in a birefringent magneto-optic material is given by~\cite{Landau}
\begin{eqnarray}
    \left(k^{2}_{0}\:\tilde{\epsilon}-
    k^2\:\mathbf{I}+\mathbf{k}\,\mathbf{k}\right)\cdot
    \,\mathbf{E_0}=0.
\end{eqnarray}
where $k_0=\omega/c$, $c$ is the speed of light in vacuum, $\mathbf{E}_0$ is the plane wave amplitude, $\mathbf{k}\,\mathbf{k}$ is a dyadic product of the wave vector, and $\mathbf{I}$ is the identity matrix. At optical wavelengths, the relative permeability $\mu$ differs only slightly from unity whereas the relative permittivity tensor $\tilde{\epsilon}$ has the form
\begin{equation}
\label{1}
    \tilde{\epsilon}=\left(
    \begin{array}{c c c}
    \epsilon_{xx} & i\epsilon_{xy} & 0\\
    -i\epsilon_{xy} & \epsilon_{yy} & 0\\
    0 & 0 & \epsilon_{zz}\\
    \end{array}
    \right).
        \end{equation}
Here the saturation magnetization of the magnetic medium is directed along the $z$-axis and rotational symmetry about this axis is assumed. Linear birefringence is expressed through the difference in diagonal components $\epsilon_{xx}$ and $\epsilon_{yy}$, and gyrotropy through the off-diagonal elements $i\epsilon_{xy}$ and $-i\epsilon_{xy}$.
Our treatment assumes a monochromatic time dependence $\exp(i \omega t)$ and no optical absorption. By solving the wave equation in this birefringent magneto-optic medium, one obtains the eigenvectors
\begin{eqnarray}
 \label{3}
    \mathbf{\hat{e}}_{\pm}=\frac{1}{\sqrt{2}}
    \left(
                \begin{array}{c}
                \cos\alpha \pm \sin \alpha \\
                \pm i\cos \alpha - i \sin \alpha\\
                0\\
                \end{array}
    \right),
\end{eqnarray}
with refractive indices $n_{\pm}$, and $n^2_{\pm}=\bar{\epsilon} \pm  \sqrt{\Delta^2+\epsilon_{xy}^2}$, respectively.
Here $\bar{\epsilon}=(\epsilon_{yy} + \epsilon_{xx})/2$, $\Delta=(\epsilon_{yy} - \epsilon_{xx})/2$, and $\alpha$, referred hereafter as \emph{elliptical birefringence parameter}, defined as $\alpha=\gamma/2$, with $\gamma$ given by
\begin{eqnarray}
\label{5}
\left\{
\begin{array}{c}
  \sin \gamma=\frac{\Delta}{\sqrt{\Delta^2 + \epsilon^2_{xy}}} \\
  \cos \gamma=\frac{\epsilon_{xy}}{\sqrt{\Delta^2 + \epsilon^2_{xy}}}
  \end{array}
  \right. .
\end{eqnarray}
In deriving Eq.~(\ref{3}), wave propagation parallel to the $z$-axis is assumed (Fig.~\ref{fig1}). Expressions~(\ref{3}) for the eigenvectors are identical to the elliptical eigenmodes introduced in Ref.~\cite{Levy2006} for elliptically birefringent magneto-optic media. These eigenvectors can be simply related to circularly polarized modes through a rotation transformation on $\mathbf{\hat{c}}_{+}$ and $\mathbf{\hat{c}}_{-}$
\begin{eqnarray}
\label{6}
    \left(
                \begin{array}{c}
                \mathbf{\hat{e}}_{+} \\
                \mathbf{\hat{e}}_{-}\\
                \end{array}
    \right) &=&
    \left(
                \begin{array}{c c}
                \cos\frac{\gamma}{2} & \sin\frac{\gamma}{2} \\
                -\sin\frac{\gamma}{2} & \cos \frac{\gamma}{2}\\
                \end{array}
    \right)
       \left(
                \begin{array}{c}
                \mathbf{\hat{c}}_{+} \\
                \mathbf{\hat{c}}_{-} \\
                \end{array}
    \right),
\end{eqnarray}
where
\begin{eqnarray}
\label{7}
    \mathbf{\hat{c}}_\pm=\frac{1}{\sqrt{2}}
    \left(
                \begin{array}{c}
                1 \\
                \pm i\\
                \end{array}
        \right).
\end{eqnarray}
\section{One-dimensional birefringent magnetophotonic crystal}
\begin{figure}[t]
\includegraphics{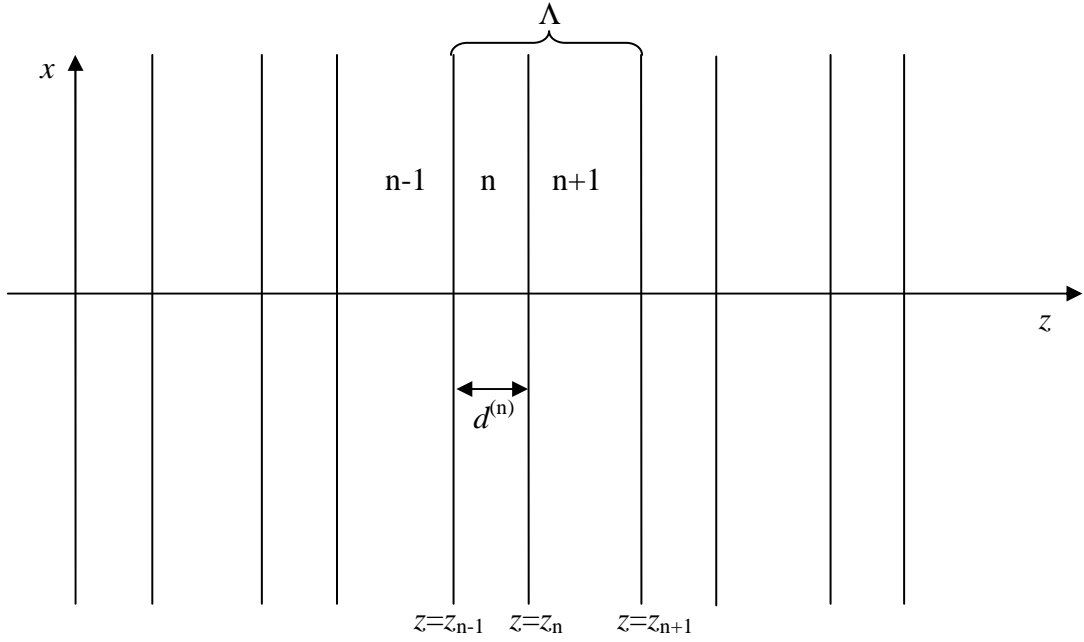}
\caption{Schematic diagram of a one-dimensional birefringent magnetophotonic crystal with period of $\Lambda$. The magnetophotonic crystal extend indefinitely in the $x$ and $y$ directions. A plane wave is incident from the left onto the structure. }
\label{fig1}
\end{figure}

The basic geometry of our model is depicted in Fig.~(\ref{fig1}). Plane waves are normally incident on a periodic stack structure consisting of alternating elliptically birefringent magneto-optic layers. The layers have different average dielectric constants $\bar \epsilon$, and are not assumed to have the same linear birefringence terms $\Delta$, or gyrotropic components $\epsilon_{xy}$. In addition, the model does not assume quarter-wave plates but allows the layers to have arbitrary thickness. Thus the model is quite general and does not impose any constraints on the relative linear birefringence, gyrotropy, or thickness of the layers. The electric field in the $n$-th layer of the periodic stack  is a linear combination of the elliptical eigenvectors obtained in the previous section.
\begin{eqnarray}
\label{8}
    \mathbf{E}(z)&=&\left[E_{01} \exp\left(-i\frac{\omega}{c}n_{+}(z-z_n)\right)+ E_{02} \exp\left(i\frac{\omega}{c}n_{+}(z-z_n)\right)\right]\mathbf{\hat{e}_{+}} \\
    \nonumber &&+\left[E_{03} \exp\left(-i\frac{\omega}{c}n_{-}(z-z_n)\right)+ E_{04} \exp\left(i\frac{\omega}{c}n_{-}(z-z_n)\right)\right]\mathbf{\hat{e}_{-}}.
\end{eqnarray}
From Eq.~(\ref{3}) we notice that it is essential to perform the transformation introduced in Eq.~(\ref{6}) to circularly polarized representation, since the eigenvectors $\mathbf{\hat{e}_{\pm}}$ change across the boundary. This concerns the continuity requirement of the tangential components of electric and magnetic fields at each interface. Following Yeh's matrix formulation~\cite{Yeh1979}, the transfer matrix which relates the four eigenmode amplitudes in the $(n-1)$-th layer to the amplitudes in the $n$-th layer is obtained as (Fig.~\ref{fig1})
\begin{eqnarray}
\label{9}
   \mathbf{T}^{(n-1,n)}=\left(\mathbf{D}^{(n-1)}\right)^{-1}\mathbf{D}^{(n)}\mathbf{P}^{(n)}.
\end{eqnarray}
where the dynamical and propagation matrices $\mathbf{D}$ and $\mathbf{P}$
 are defined as
 \begin{eqnarray}
 \label{10}
    \mathbf{D}^{(n)}&=&\left(
                \begin{array}{c c c c}
                \cos \alpha^{(n)} & \cos \alpha^{(n)}& -\sin\alpha^{(n)} & -\sin\alpha^{(n)} \\
                \cos \alpha^{(n)}n^{(n)}_{+}& -\cos \alpha^{(n)}n^{(n)}_{+}& -\sin\alpha^{(n)} n^{(n)}_{-}& \sin\alpha^{(n)} n^{(n)}_{-} \\
                \sin\alpha^{(n)} & \sin\alpha^{(n)} & \cos \alpha^{(n)} & \cos \alpha^{(n)} \\
                \sin\alpha^{(n)} n^{(n)}_{+} & -\sin\alpha^{(n)} n^{(n)}_{+} & \cos \alpha^{(n)}n^{(n)}_{-} & -\cos \alpha^{(n)}n^{(n)}_{-}
                \end{array}
    \right),
    \end{eqnarray}
 \begin{eqnarray}
 \label{11}
  \mathbf{P}^{(n)}&=&\left(
                \begin{array}{c c c c}
                e^{i \beta^{(n)}_{+}}& 0 & 0 & 0\\
                0& e^{-i \beta^{(n)}_{+}} & 0 & 0\\
                0& 0 & e^{i \beta^{(n)}_{-}} & 0\\
                0& 0 & 0 & e^{-i \beta^{(n)}_{-}}
                \end{array}
        \right),
\end{eqnarray}
with $\beta^{(n)}_{\pm}=(\omega/c)n^{(n)}_{\pm}d^{(n)}$, where $d^{(n)}$ is the thickness of the $n$-th layer.

From Eqs.~(\ref{9})-(\ref{11}) we note that for zero birefringence the transfer matrix becomes block diagonal and reproduces the results obtained in Ref.~\cite{Visnovsky2001a} for circularly polarized eigenmodes.
\section{Bloch modes and dispersion relation}
Consider a unit cell of the birefringent magnetophotonic crystal. This consists of two adjacent layers as shown in Fig.~\ref{fig1}. According to the Floquet-Bloch theorem, the optical electric field obeys the relation
\begin{eqnarray}
\mathbf{E}(z)=\mathbf{E}_K(z) \exp(i K z),
\end{eqnarray}
where $\mathbf{E}_K(z)$ is a periodic function of coordinate $z$ with a period $\Lambda$.

From Eq.~(\ref{9}) the transfer matrix for a unit cell of the crystal obeys
\begin{eqnarray}
 \label{12}
 \mathbf{T}^{(n-1,n+1)}=\left(\mathbf{D}^{(n-1)}\right)^{-1} \underbrace{\mathbf{D}^{(n)}\mathbf{P}^{(n)}
 \left(\mathbf{D}^{(n)}\right)^{-1}}_{\mathbf{S}^{(n)}}\mathbf{D}^{(n+1)}\mathbf{P}^{(n+1)}.
\end{eqnarray}
The Floquet-Bloch theorem thus imposes the following condition:\cite{comment2}
\begin{eqnarray}
\label{13}
  \mathbf{T}^{(n-1,n+1)}\mathbf{E} &=& \lambda \mathbf{E}.
\end{eqnarray}
To solve this eigenvalue problem we note that the term $\mathbf{S}^{(n)}$ in Eq.~(\ref{12}) can be transformed into block diagonal form through a similarity matrix transformation as follows
\begin{eqnarray}
\label{14}
  \mathbf{S}^{(n)} &=& \left(\mathbf{U}^{(n)}\right)^{-1}\mathbf{S}^{(n)}_{c}\mathbf{U}^{(n)}.
\end{eqnarray}
where $\mathbf{U}^{(n)}$ and $\mathbf{S}^{(n)}_{c}$ are given by
\begin{eqnarray}
    \mathbf{U}^{(n)}=\left(
    \begin{array}{c c c c}
    \cos\alpha^{(n)} & 0 & \sin\alpha^{(n)} & 0 \\
    0 & \cos\alpha^{(n)} & 0 & \sin\alpha^{(n)} \\
    -\sin\alpha^{(n)} & 0 & \cos\alpha^{(n)} & 0 \\
    0 & -\sin\alpha^{(n)} & 0 & \cos\alpha^{(n)} \\
    \end{array}
    \right)
\end{eqnarray}
and
\begin{eqnarray}
\mathbf{S}^{(n)}_{c}=
\begin{pmatrix}
  \cos \beta^{(n)}_+ & \frac{i}{n^{(n)}_+}\sin \beta^{(n)}_+ & 0 & 0 \\
  i n^{(n)}_+\sin \beta^{(n)}_+  & \cos \beta^{(n)}_+ & 0 & 0 \\
  0 & 0 & \cos \beta^{(n)}_- & \frac{i}{n^{(n)}_{-}}\sin \beta^{(n)}_{-} \\
  0 & 0 & i n^{(n)}_{-}\sin \beta^{(n)}_{-} & \cos \beta^{(n)}_{-} \\
\end{pmatrix},
\end{eqnarray}
respectively.
Upon substitution of Eq.(\ref{14}) into Eq.(\ref{12}), we obtain
\begin{eqnarray}
\label{15}
  \mathbf{T}^{(n-1,n+1)}=\left(\mathbf{\Phi}^{(n,n+1)}\right)^{-1} \mathbf{S}^{(n)}_{c}\mathbf{\Phi}^{(n,n+1)}\mathbf{P}^{(n+1)},
\end{eqnarray}
with
\begin{eqnarray}
\label{16}
  \mathbf{\Phi}^{(n,n+1)} &=& \mathbf{U}^{(n)}\mathbf{D}^{(n+1)}.
\end{eqnarray}
The matrix $\mathbf{\Phi}^{(n,n+1)}$ is given by
\begin{eqnarray}
\label{17}
  \begin{pmatrix}
    \cos \chi^{(n,n+1)} & \cos \chi^{(n,n+1)} & \sin \chi^{(n,n+1)} & \sin \chi^{(n,n+1)} \\
    n^{(n+1)}_{+} \cos \chi^{(n,n+1)} & -n^{(n+1)}_{+} \cos \chi^{(n,n+1)} & n^{(n+1)}_{-} \sin \chi^{(n,n+1)} & -n^{(n+1)}_{-} \sin \chi^{(n,n+1)} \\
    -\sin \chi^{(n,n+1)} & -\sin \chi^{(n,n+1)} & \cos \chi^{(n,n+1)} & \cos \chi^{(n,n+1)} \\
     -n^{(n+1)}_{+} \sin \chi^{(n,n+1)} & n^{(n+1)}_{+} \sin \chi^{(n,n+1)} & n^{(n+1)}_{-} \cos \chi^{(n,n+1)} & -n^{(n+1)}_{-} \cos \chi^{(n,n+1)} \\
  \end{pmatrix},
\end{eqnarray}
where $\chi^{(n,n+1)}=\alpha^{(n)}-\alpha^{(n+1)}$. We notice that only the relative elliptical birefringence parameter $\chi^{(n,n+1)}$ appears in the unit cell transformation matrix.

\begin{figure}[t]
\includegraphics{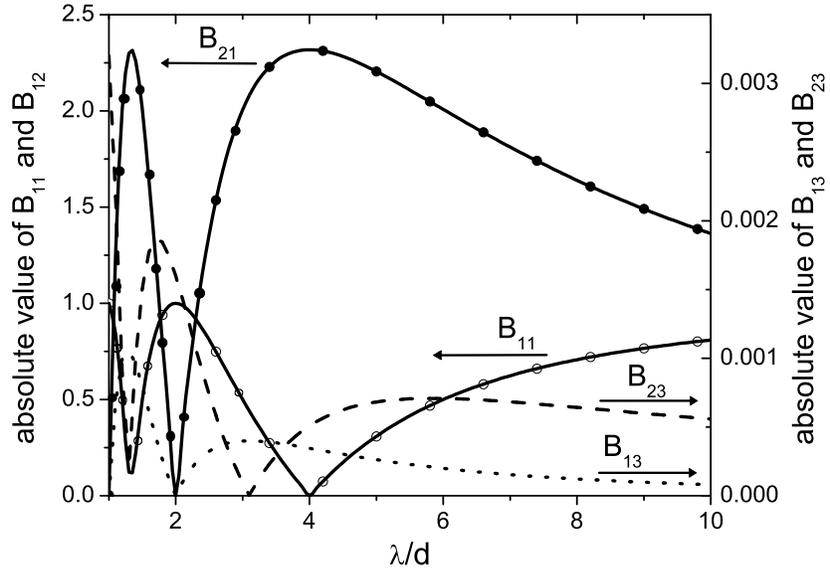}
\caption{Variation of elements of the $\mathbf{B}$ matrix versus layer thickness $d^{(n)}$ in the unit cell for a typical [Bi,Lu]$_3$Fe$_5$O$_{12}$ layer. Solid lines with circles and dots are curves for  $B_{11}$ and $B_{21}$ from block-diagonal elements, respectively. Dashed and dotted lines are curves for $B_{23}$ and $B_{13}$ from off-block diagonal elements, respectively. }
\label{fig2}
\end{figure}

The advantage of reformulating the problem in this way is that it allows us to express the Floquet-Bloch condition in terms of block diagonal matrices, making it possible to obtain fully analytic solutions. Upon substitution of Eq.~(\ref{15}) into Eq.~(\ref{13}) we obtain a generalized eigenvalue equation of the form
\begin{eqnarray}
\label{eig}
   \mathbf{S}^{(n)}_{c} \mathcal{E}&=& \lambda \mathbf{B} \mathcal{E},
\end{eqnarray}
 where $\mathcal{E}=\mathbf{\Phi}^{(n,n+1)}\mathbf{P}^{n+1} \mathbf{E}$. The eigenvalues of the transformation matrix $\mathbf{T}^{(n-1,n+1)}$ are given by the solutions to
\begin{eqnarray}
\label{18}
  \left|\mathbf{S}^{(n)}_{c}-\lambda \mathbf{B}\right| &=& 0.
\end{eqnarray}
The elements of the matrix $\mathbf{B}$ are given by
\begin{eqnarray}
B_{1,1}&=& B_{2,2}=\cos^{2}\chi^{(n,n+1)} \cos{\beta^{(n+1)}_{+}} +  \sin^{2}\chi^{(n,n+1)} \cos{\beta^{(n+1)}_{-}},\\
B_{1,2}&=&-i \frac{\sin^{2}\chi^{(n,n+1)}  \sin{\beta^{(n+1)}_{-}}}{n^{(n+1)}_{-}}-i \frac{\cos^{2}\chi^{(n,n+1)} \sin{{\beta^{(n+1)}_{+}}}}{n_{+}^{(n+1)}},\\
B_{1,3}&=&B_{3,1}=B_{2,4}=B_{4,2}= \frac{1}{2}\sin2\chi^{(n,n+1)}\left(\cos{{\beta^{(n+1)}_{-}}}-\cos{{\beta^{(n+1)}_{+}}}\right)\approx 0, \\
B_{1,4}&=&B_{3,2}=\frac{i \sin2\chi^{(n,n+1)}}{2 \;n^{(n+1)}_{-} n^{(n+1)}_{+}} \left(n^{(n+1)}_{-} \sin \beta^{(n+1)}_{+} - n^{(n+1)}_{+} \sin \beta^{(n+1)}_{-} \right)\approx 0,\\
B_{2,1}&=& -i n^{(n+1)}_{-} \sin^{2}\chi^{(n,n+1)} \sin{\beta^{(n+1)}_{-}}-i n_{+}^{(n+1)}\cos^{2}\chi^{(n,n+1)} \sin{{\beta^{(n+1)}_{+}}}, \\
B_{2,3}&=&B_{4,1}= i\; \frac{1}{2}\sin2\chi^{(n,n+1)}\left(n^{(n+1)}_{+} \sin \beta^{(n+1)}_{+} -  n^{(n+1)}_{-} \sin \beta^{(n+1)}_{-}\right)\approx 0,\\
B_{3,3}&=&B_{4,4}= \cos^{2}\chi^{(n,n+1)}  \cos{\beta^{(n+1)}_{-}} + \sin^{2}\chi^{(n,n+1)}  \cos{\beta^{(n+1)}_{+}},\\
B_{3,4} &=& -i \frac{\cos^{2}\chi^{(n,n+1)}  \sin{\beta^{(n+1)}_{-}}}{n^{(n+1)}_{-}}-i \frac{\sin^{2}\chi^{(n,n+1)} \sin{{\beta^{(n+1)}_{+}}}}{n_{+}^{(n+1)}},\\
B_{4,3}&=& -i n^{(n+1)}_{-} \cos^{2}\chi^{(n,n+1)}  \sin{\beta^{(n+1)}_{-}}-i n_{+}^{(n+1)} \sin^{2}\chi^{(n,n+1)} \sin{{\beta^{(n+1)}_{+}}}.
\end{eqnarray}
Notice that all off-block diagonal components contain a difference term between trigonometric functions of $\beta_{+}$ and $\beta_{-}$. For all available magnetic garnet materials in the infrared wavelength range, and most other Faraday rotation materials of interest such as Fe:InGaAsP and CdMnHgTe~\cite{Zaman2007, Hwang2006}, this difference is very small. For example, a typical value for the dielectric tensor of bismuth-substituted lutetium iron garnet is~\cite{Levy2006}
\begin{equation}
\label{20}
    \begin{pmatrix}
    5.369 & i 0.00274 & 0\\
    -i 0.00274 & 5.373 & 0\\
    0 & 0 & \epsilon_{zz}\\
    \end{pmatrix},
\end{equation}
yielding, for one-quarter and one-third wave plates, the following $\mathbf{B}$ matrices
\begin{eqnarray}
\left(
\begin{array}{llll}
 -0.000681 & - i 0.431 & -0.000196 & i 0.0000539  \\
 - i 2.32  & -0.000681 & - i 0.00029  & -0.000196 \\
 -0.000196 & i 0.0000539 & 0.000681 & - i 0.432 \\
 - i 0.00029 & -0.000196 & - i 2.32 & 0.000681
\end{array}
\right)
\end{eqnarray}
and
\begin{eqnarray}
\left(
\begin{array}{llll}
 -0.501 & - i 0.373  & -0.000227 &  i 0.000103 \\
 - i 2.01 & -0.501 &  i 0.0000525 & -0.000227 \\
 -0.000227 & i 0.000103 & -0.499 & - i 0.374  \\
 i 0.0000525  & -0.000227 & - i 2.01  & -0.499
\end{array}
\right),
\end{eqnarray}
respectively. In both cases we can see the off-block diagonal elements to be much smaller than the block diagonal elements. This is also true for technologically important non-garnet materials such as Fe:InGaAsP with Verdet coefficient of $12.5^{\circ}$~/~mm~T and $\bar \epsilon=9.61$ at a wavelength of $1.55~\mu m$, where similar order magnitude values for the $\mathbf{B}$ matrix are obtained.~\cite{Zaman2007} In Fig.~\ref{fig2} we show the variation of four elements of the $\mathbf{B}$ matrix, two elements from the block diagonal and two off-block diagonal elements, versus layer thickness $d^{(n)}$ assuming the dielectric tensor Eq.~(\ref{20}). For any value of layer thickness ($d^{(n)}<$ wavelength) the off-block diagonal terms are small.

Moreover, all off-block diagonal elements, themselves already small, will be raised to the second or fourth powers in the determinant of the $4\times4$ matrix making their contributions even smaller. Therefore, we can safely ignore the effect of  off-block diagonal terms on the determinant and set them equal to zero. The problem now reduces to an eigenvalue problem for two $2 \times 2$  matrices. The solution for the $\lambda$'s yields the Bloch phase factor $\exp( i K \Lambda)$ and the dispersion relation,
\begin{eqnarray}
\label{19}
    \cos K \Lambda &=&  B_{\pm}\cos \beta^{(n)}_{\pm} -\frac{1}{2} C_{\pm} \sin \beta^{(n)}_{\pm},
  \end{eqnarray}
  with $B_{+}(B_{-})= B_{1,1}(B_{3,3})$. The factors $C_{\pm}$ in Eq.~(\ref{19}) are defined as
\begin{eqnarray}
C_{\pm}&=& i n^{(n)}_{\pm}B_{i,j}+ \frac{i}{n^{(n)}_{\pm}}B_{j,i},
\end{eqnarray}
with the subscripts $\{i,j\}$ take $\{1,2\}(\{3,4\})$ for $\{+\}(\{-\})$ sign. Equation~(\ref{19}) determines the dispersion relation for the Bloch wave vector $K$ with  frequency.

If the linear birefringence term $\Delta$ in this expression is set equal to zero, and the unit cell is composed of a thin magneto-optic layer and an isotropic layer, Eq.~(\ref{19}) reduces to the ultra-thin approximation of Ref.~\cite{Khanikaev2005} upon letting the thickness of the magneto-optic layer tend to zero. In this limiting case Eq.~(\ref{19}) becomes:
\begin{eqnarray}
\cos K \Lambda \approx \cos k_0 \Lambda -\frac{1}{2}\left(m \pm \frac{1}{2} \Delta\right) k_0 \sin k_0 \Lambda,
\end{eqnarray}
in agreement with Ref.~\cite{Khanikaev2005}. Here $\Delta=\left(n^{2}_{+}-n^{2}_{-}\right)d$ and $m=\bar n^2 d$ are parameters used in the above referenced work.
The present formulation also agrees with the treatment in Ref.~\cite{Levy2006} that discusses the limit of uniform linear birefringence across the photonic crystal and where the difference between refractive indices of the two layers in the unit cell is small. In that case $\alpha^{(n)}=\alpha^{(n+1)}$ and dispersion relation Eq.~(\ref{19}) reduces to
\begin{eqnarray}
\label{22}
  \cos K \Lambda &=& \cos\left(\beta^{(n)}_{\pm}+\beta^{(n+1)}_{\pm}\right),
  \end{eqnarray}
  yielding
  \begin{eqnarray}
  K=\frac{\omega}{c}\left(\frac{n^{(n)}_{\pm} d^{(n)}+n^{(n+1)}_{\pm} d^{(n+1)}}{d^{(n)}+d^{(n+1)}}\right)\equiv \frac{\omega}{c} \bar n_{\pm}
\end{eqnarray}
with identical elliptical eigenmodes and average indices $\bar n_{\pm}$ to those in Ref.~\cite{Levy2006}.

The eigenvectors in Eq.~(\ref{13}) are
 \begin{eqnarray}
 \label{21}
 \mathbf{E}=\left(\mathbf{\Phi}\mathbf{P}^{(n+1)}\right)^{-1}
    \begin{pmatrix}
     \frac{\frac{-i}{n^{(n)}_{+}}\sin \beta^{(n)}_{+}+B_{1,2}e^{\pm i K_{\pm} \Lambda} }{\cos \beta^{(n)}_{+}-B_{+}e^{\pm i K_{\pm} \Lambda}} \\
     1\\
     \frac{\frac{-i}{n^{(n)}_{-}}\sin \beta^{(n)}_{-}+B_{3,4}e^{\pm i K_{\pm} \Lambda }}{\cos \beta^{(n)}_{-}-B_{-}e^{\pm i K_{\pm} \Lambda}} \\
     1\\
     \end{pmatrix}
 \end{eqnarray}
 This model yields elliptically polarized solutions with different elipticity across layer boundaries, according to Eq.~(\ref{8}). From the eigenmodes Eq.~(\ref{21}) we notice that the elliptically polarized electric field of a layer in a unit cell not only depends on the elliptical birefringence parameter $\alpha^{(n)}$ of that layer but also on the relative elliptical birefringence parameter $\chi^{(n,n+1)}$.

\begin{figure}[t]
\includegraphics{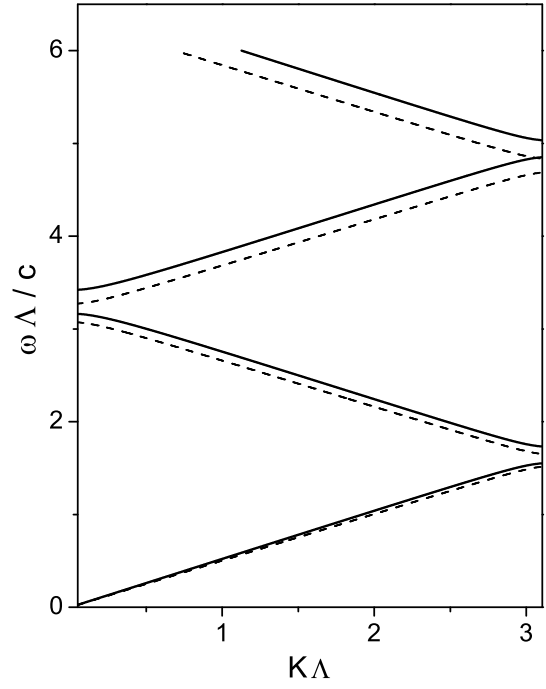}
\caption{Band structure for a periodic stack of Bi:LuIG and LaGG  with $d^{(n)}=0.3 \Lambda$ and $d^{(n+1)}=0.7 \Lambda$, respectively. The dashed line shows the $K_{+}$ wave branch and the black line shows the $K_{-}$ wave branch. }
\label{fig3}
\end{figure}

\begin{figure}[t]
\includegraphics{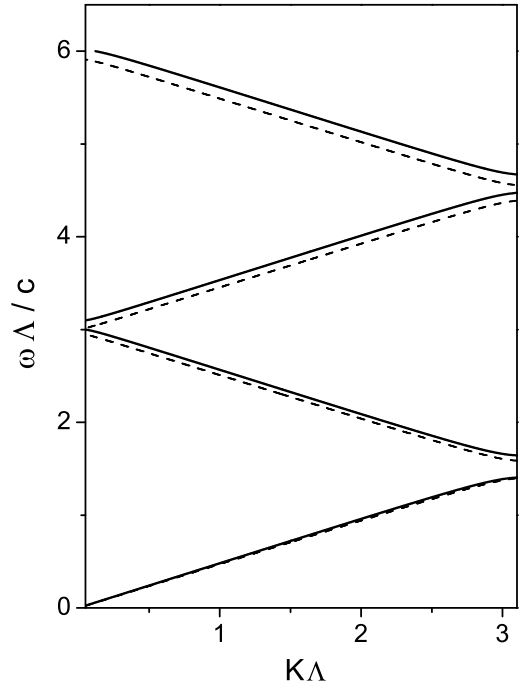}
\caption{Band structure for a periodic stack of Bi:LuIG and LaGG  with $d^{(n)}=d^{(n+1)}=0.5 \Lambda$. The dashed line shows the $K_{+}$ wave branch and the black line shows the $K_{-}$ wave branch. }
\label{fig4}
\end{figure}

 Band structures based on dispersion relation Eq.~(\ref{19}) for two crystals having different unit cells are plotted in Figs.~(\ref{fig3}) and (\ref{fig4}). The unit cell consists of one birefringent [Bi,Lu]$_3$Fe$_5$O$_{12}$ layer adjacent to a Lanthanum garnet layer (La$_3$Ga$_5$O$_{12}$) with the dielectric constant given in Ref.~\cite{Khartsev2005c}. Figure~(\ref{fig3}) corresponds to the case $d^{(n)}=d^{(n+1)}=0.5 \Lambda$, and Fig.~(\ref{fig4}) to $d^{(n)}=0.3 \Lambda$.

 Although the gyrotropic components $\epsilon_{xy}$ and linear birefringence term $\Delta$ are naturally small we are able to tell apart the two branches of the electromagnetic modes in the unit cell especially in the higher Bragg orders. Of particular note is the fact that the relative thickness of layers $d^{(n)}$ and $d^{(n+1)}$ strongly affects the band structure, as can be seen by comparing the third and fourth-order Bragg mode branches in Fig.~\ref{fig3} and Fig.~\ref{fig4}.  Although the material compositions are the same in both figures, a departure from the quarter-wave plate configuration in Fig.~\ref{fig3} shifts the band gaps enough to induce a cross-over between different helicity modes in the high-order branches.We have also calculated numerically the band structures using the full form of the $\mathbf{B}$ matrix in eigenvalue Eq.~(\ref{eig}) and compared with the analytic expression Eq.~(\ref{19}). The maximum difference between the two formulations is at most 0.3\%.

\section{Conclusions}
We have presented a fully analytical treatment of the band structure and Bloch states for one-dimensional elliptically birefringent magnetophotonic crystals.  Expressions for the dispersion relation and Bloch modes are obtained.  This formulation is applicable to most technologically important magneto-optic systems such as iron garnets in the near infrared and diluted magnetic semiconductors in high and low magnetic fields.  The model treats the case of photonic crystals in magnetic systems in the presence of non-circular birefringence, such as result from strains due to lattice mismatch and differential thermal expansion with the growth substrate.  In other words, the formulation presented here accounts for gyrotropy and linear birefringence simultaneously.  This is particularly noteworthy because of the large class of photonic bandgap structures on magneto-optic films encompassed under this category and that naturally arise from standard fabrication techniques.

A key feature of the treatment is the reduction of the Floquet-Bloch eigenvalue equation to block diagonal form, and hence its analytic solution.  The model reduces to the exact solution for one-dimensional magnetophotonic crystals with ultrathin magnetic layers in the absence of linear birefringence.  It also contains the solution for magnetophotonic crystals with uniform birefringence as a limiting case.  Both of these models have been recently discussed in the literature.\cite{Khanikaev2005,Levy2006} The power of the formulation presented here is that it applies to layers of arbitrary thickness and elliptical birefringence level and that it allows for alternating birefringence from one layer to the next.

Band structures and the form of the Bloch wavefunctions are found to depend explicitly on the relative birefringence parameter $\chi^{(n,n+1)}$ .  Different branches arise in the band structure separating Bloch modes with positive and negative helicity.  It is seen that the frequency shift between these branches is strongly dependent on the relative thickness of adjacent layers in the stack.  The band gaps grow with Bragg mode number generating cross-over points between different helicity states in high order for asymmetric structures.

\section*{Acknowledgments}
This material is based upon work supported by the National Science Foundation under Grant No. ECCS-0520814.


\begin{thebibliography}{23}
\bibitem{comment}
   The use of TE/TM terminology in this context is meant to illustrate the fact that different waveguide modes with different polarization states possess different modal refractive indices.  The authors are aware that the polarization state of light in gyrotropic films, especially under material stresses and the geometrical confinement generated by asymmetric optical waveguide channels should be described in terms of waveguide modes that account for elliptical birefringence. A full treatment of the polarization state of Bloch modes is presented in this work.
\bibitem{comment2}
We use the greek letter $\lambda$ represent the eigenvalue of the Block equation. It is not used as a symbol for wavelength.
\end{thebibliography}

\newpage

%
%
%
%
%

\end{document}